\documentclass[aip,jap,reprint, amsmath,amsfonts,amssymb]{revtex4-1}
\usepackage{rotating}
\usepackage{graphicx}

\def\KTH{Nanostructure Physics, Royal Institute of Technology (KTH), Albanova, SE-10791 Stockholm, Sweden}
\def\SU{Department of Physics, Stockholm University, 106 91 Stockholm, Sweden}
\def\NTNU{NTNU Norwegian University of Science and Technology, Dept. of Engineering Design and Materials, 7491 Trondheim, Norway}

\begin{document}

\title{Imaging high-speed friction at the nanometer scale}

\author{Per-Anders Thor\'{e}n*}
\email{pathoren@kth.se}
\affiliation{\KTH}
\author{Astrid S. de Wijn}
\email{dewijn@fysik.su.se,astrid@dewijn.eu}
\affiliation{\SU; \NTNU}
\author{Riccardo Borgani}
\affiliation{\KTH}
\author{Daniel Forchheimer}
\affiliation{\KTH}

\author{David B. Haviland}
\email{haviland@kth.se}
\affiliation{\KTH}

\maketitle

\textbf{Friction is a complicated phenomenon involving nonlinear dynamics at different length and time scales~\cite{Vanossi2013,perssonbook}.
The microscopic origin of friction is poorly understood, due in part to a lack of methods for measuring the force on a nanometer-scale asperity sliding at velocity of the order of cm/s.~\cite{krimreview,Holscher2008}  Despite enormous advance in experimental techniques~\cite{meyerbook}, this combination of small length scale and high velocity remained illusive.  Here we present a technique for rapidly measuring the frictional forces on a single asperity (an AFM tip) over a velocity range from zero to several cm/s.  At each image pixel we  obtain the velocity dependence of both conservative and dissipative forces, revealing the transition from stick-slip to a smooth sliding friction \cite{Vanossi2013, stickslip_smoothsliding}.  We explain measurements on graphite using a modified Prandtl-Tomlinson model that takes into account the damped elastic deformation of the asperity. With its greatly improved force sensitivity and very small sliding amplitude, our method enables rapid and detailed surface mapping of the full velocity-dependence of frictional forces with less than 10~nm spatial resolution. }

\begin{figure}
\includegraphics[width=1.0\columnwidth]{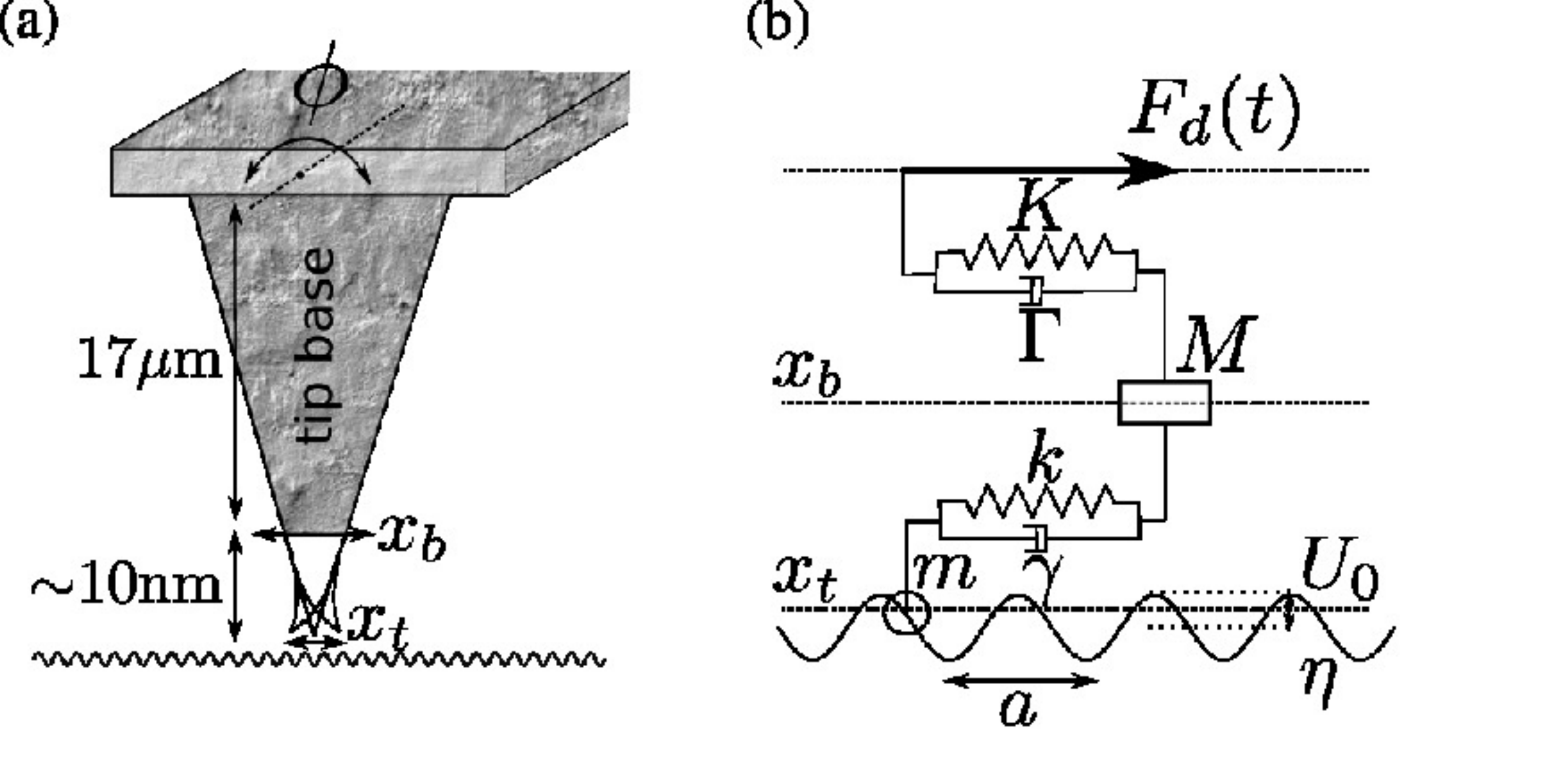}
\caption{
\textbf{A schematic of the experiment and model, not to scale.}  a) The AFM cantilever undergoes a twisting oscillation at the resonance frequency of a high-Q torsional eigenmode.  The resulting lateral motion of the tip base $x_\mathrm{b}$ is dampened by frictional forces acting on the tip apex, $x_\mathrm{t}$.
 b) Schematic of the modified Prandtl-Tomlinson (PT) model use to describe the dynamical system.  A driven support (cantilever base) is coupled to the nonlinear surface potential via a linear oscillator (torsional resonance) and elastic asperity (tip).   
}
\label{fig:twist}
\end{figure}

Many applications in tribology require an understanding of frictional forces on nanometer-scale contacts with a relative velocity of at least 1 cm/s.  Traditional measurement of nanoscale friction scans an Atomic Force Microscope (AFM) tip or colloidal probe across a surface at constant velocity \cite{Qian2013,Alvarez-Asencio2013}.  Friction induces a lateral force on the tip, resulting in a twist $\phi$ around the major axis of the cantilever, detected by optical beam deflection (see fig.~\ref{fig:twist}).  The cantilever's restoring force is assumed to be in quasi-static equilibrium with the lateral force on the tip, and measurement of cantilever twisting is limited by detector noise.  Unity signal-to-noise ratio in a 1 ms measurement time defines a minimum detectable force $F_\text{min} \sim 13$~pN (see Methods).  With this quasi-static method stick-slip behavior can be observed \cite{stickslip_smoothsliding}, but only up to velocities $\sim 10$~nm/s, at least 6 orders of magnitude below the velocity scale relevant to applications.  At high scan velocities stick-slip can not be resolved, only the mean force of sliding fricion.  Scan velocities as high as  580 $\mu$m/s have been reached \cite{fast_quasistatic}, but a measurement time of 1 ms would limit spatial resolution to 580~nm.

The relevant velocity scale is achieved with greatly improved force sensitivity by dynamic measurement of friction force, when the cantilever undergoes high frequency oscillation near a high quality factor resonance.  Dynamic friction has been probed by perturbing a flexural \cite{Gysin2011} and torsional \cite{Reinstadtler2003, Bhushan2004, Yurtsever2009} resonance, but thus far these methods have not measured force, only changes of oscillation amplitude and phase when the tip engages a surface.  Near torsional resonance a good detector can measure the twisting Brownian motion of the cantilever, meaning that the minimum detectable force is at the thermal limit.  For our cantilever  $F_\text{min} = 0.88$~pN in the same 1ms measurement time (see Methods).  The high frequency of a stiff torsional resonance $\sim 2$~MHz allows for tip velocity $v_\text{max}\sim 6$~cm/s with very small amplitude of sliding oscillation $A \sim 5$~nm. 

In this article we describe calibrated and quantitative measurement of dynamic frictional force, both the conservative force $F_I$ and dissipative force $F_Q$, arising from a single asperity (the AFM tip) rapidly sliding on the surface.  We observe the transition from stick-slip to smooth sliding friction as a characteristic shape in the amplitude dependence of the dynamic force quadratures $F_I(A)$ and $F_Q(A)$.   We scan at normal speed for dynamic AFM while measuring this transition at each image pixel, thus creating an image with spatial resolution limited only by the amplitude of sliding motion $\sim$10~nm.  Our work extends the Intermodulation AFM method previously demonstrated for normal tip-surface force \cite{Platz2008, Platz2012} to lateral forces, important for understanding friction.  

\section*{Experiment}

Intermodulation AFM is based on the detection of high-order frequency mixing near mechanical resonance.  In this work the lowest torsional eigenmode (a linear oscillator) is driven at two frequencies near resonance.  When perturbed by the nonlinear frictional force, the resonator responds with a frequency comb of intermodulation products of the two drive tones \cite{Platz2008}. In the time domain this frequency comb corresponds to a rapid oscillation with a slowly modulated amplitude and phase.  Extracting the modulation phase allows us to resolve two components of the force, one which is in phase with the rapidly oscillating motion, and its quadrature.  These two components of the force can be plotted as functions of the slowly varying amplitude $A$ \cite{Platz2013B}.  Thus, the amplitude-dependent dynamic force quadrature $F_I(A)$ is the integrated conservative force in phase with the cantilever motion, and $F_Q(A)$ the dissipative force, in phase with the velocity (see eqs.~\ref{eq:fi} and \ref{eq:fq} in Methods).  The transition from stick-slip to free-sliding dynamics of the AFM tip is revealed by a characteristic shape of these two force quadrature curves.

\begin{figure}
\includegraphics[width=\columnwidth]{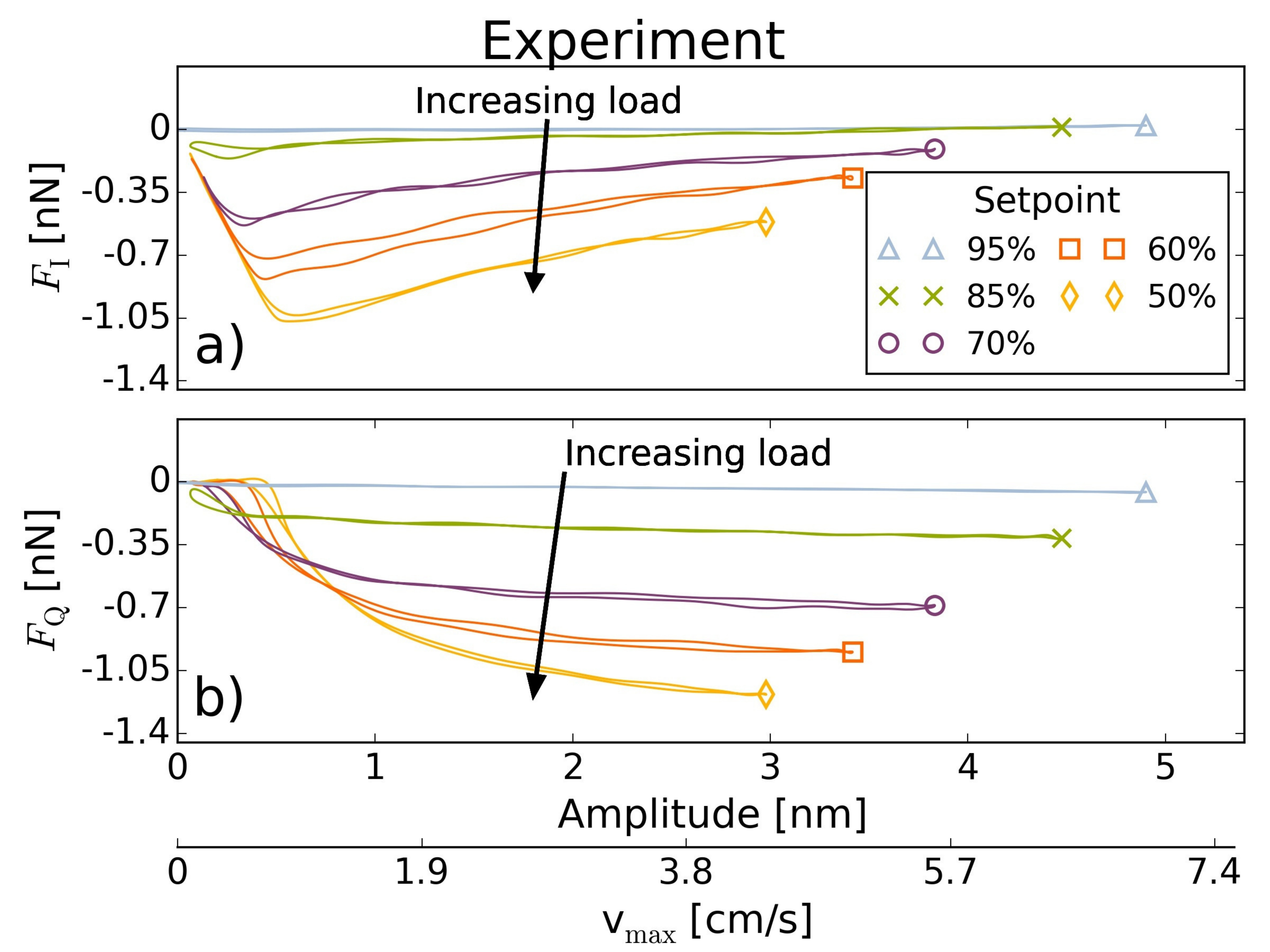}
\includegraphics[width=\columnwidth]{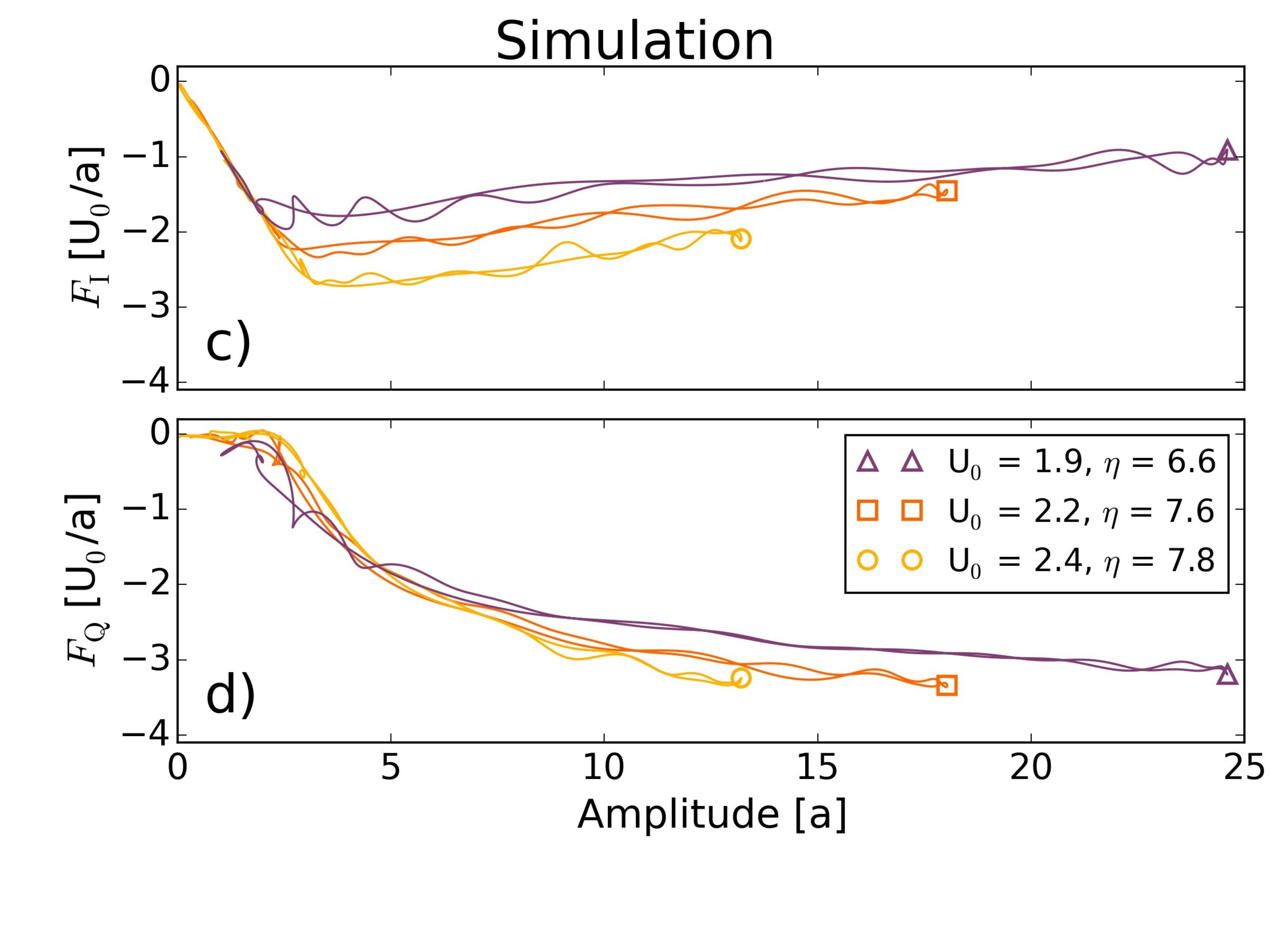}
\caption{ \textbf{Force quadratures.} a) and b) experimental curves at different probe heights (feedback set-points) showing continuous evolution from zero to increasing load.  With sufficient load force, at low-amplitude the tip is stuck and the slope of $F_I(A)$ gives the tip stiffness.   At higher amplitude stick-slip behavior gives way to smooth sliding, with $F_I$ decreasing toward zero and $F_Q$ approaching a constant value.  Qualitatively similar behaviour is seen in the simulated force quadrature curves c) and d), derived from numerical integration of a modified Prandtl-Tomlinson model. 
\label{fig:tap}
}
\end{figure}

Figure \ref{fig:tap}~a) and b) show the measured force quadrature curves for a graphite surface at different interaction strengths, realized in the experiment by changing the amplitude setpoint, which moves the AFM probe closer to the surface. At each interaction strength, the double curves show measurement with increasing and decreasing amplitude.  The net interaction which loads the frictional contact is the sum of the adhesive forces and the cantilever bending force.  The latter could in principle be measured by monitoring the vertical deflection of the cantilever.  However, with the rather stiff cantilever used in this experiment we could barely resolve a change in static bending. Adhesive forces cause 'jump-to-contact' instabilities with softer cantilevers, making it very difficult to continuously regulate the load force. In our experiment we are able to smoothly regulate the load to observe a gradual evolution of the force quadrature curves, from zero to sufficiently large interaction where linear $F_I(A)$ is observed at low critical amplitude.  

From simulations (see Theory section), this low-amplitude linear dependence of $F_I(A)$ corresponds to the tip apex being stuck to the surface. The measured cantilever motion is the result of elastic tip deformation.  At higher amplitude stick-slip dynamics begins and one observes a transition to smooth sliding with increasing oscillation amplitude, characterized by decreasing $F_I(A)$ and asymptotic approach of $F_Q(A)$ to a constant value.  One can see how reducing the interaction force results in the gradual disappearance of the low-amplitude sticking regime.  The horizontal scale of fig.~\ref{fig:tap} a) and b) also shows the maximum velocity of the tip base relative to the surface, $v_\text{max} = 2 \pi A f_0$. Note that in contrast to traditional AFM friction measurements, the cantilever is not sliding at constant velocity, but rather undergoing rapid harmonic oscillation with velocity varying between zero at each turning point, to a maximum when it crosses its torsional equilibrium point.  

\section*{Theory}

Our interpretation of the measured force quadrature curves in terms of stick-slip dynamics of a damped elastic asperity, is based on comparison of the measured data with numerical simulation of a modified Prandtl-Tomlinson (PT) model \cite{Vanossi2013, Reimann2004, Tshiprut2008, Krylov2006, highspeedsimulations, Conley2005}.
In our model (see figure~\ref{fig:twist}b) the particle is coupled via a spring and damper (damped elastic tip apex) to an intermediate support (rigid base of the tip), which in turn is coupled via a linear oscillator (cantilever torsional resonance) to a driven support (cantilever base).  The inclusion of a damped elastic tip was necessary to explain the experimental data.  

\begin{figure}
\includegraphics[width=\columnwidth]{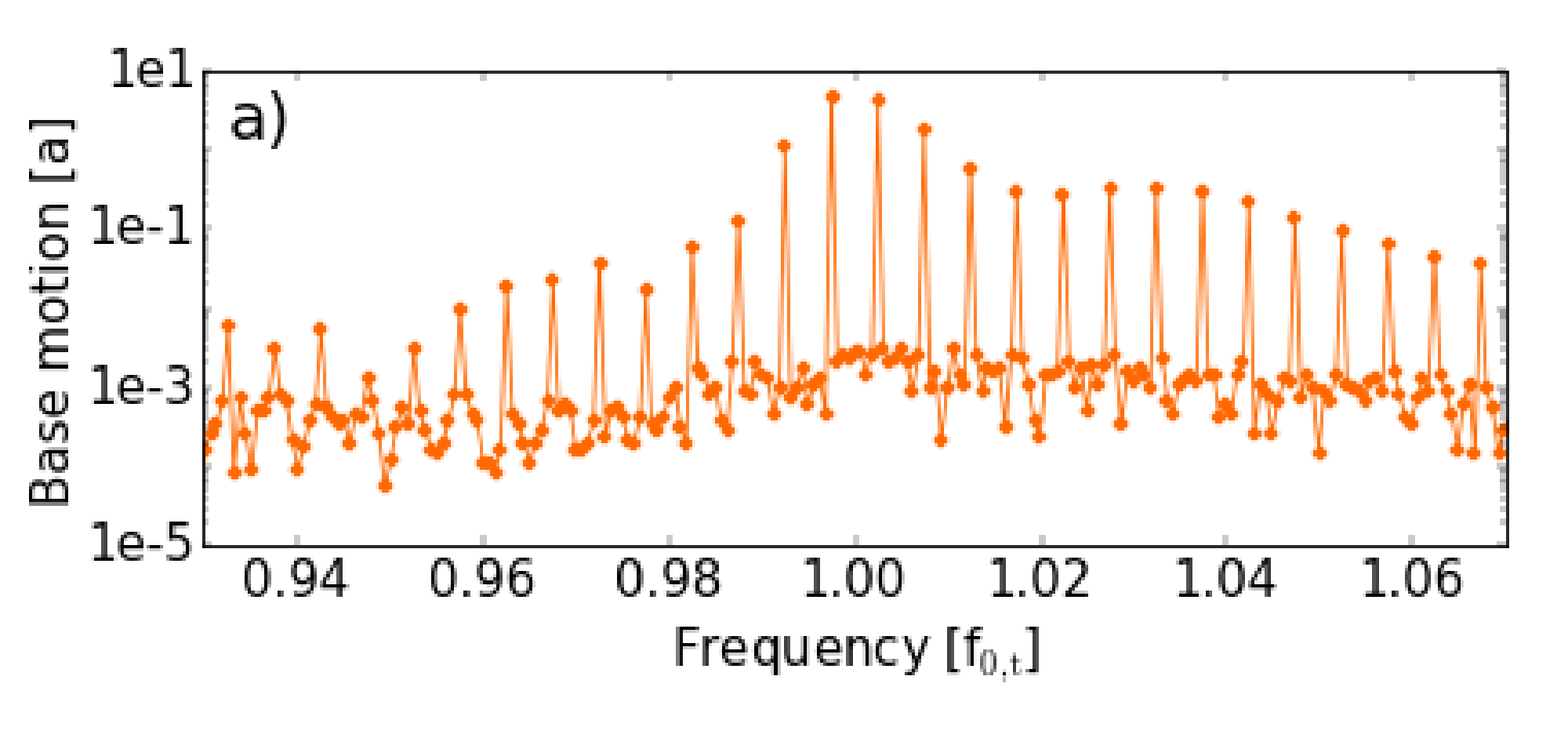}
\includegraphics[width=\columnwidth]{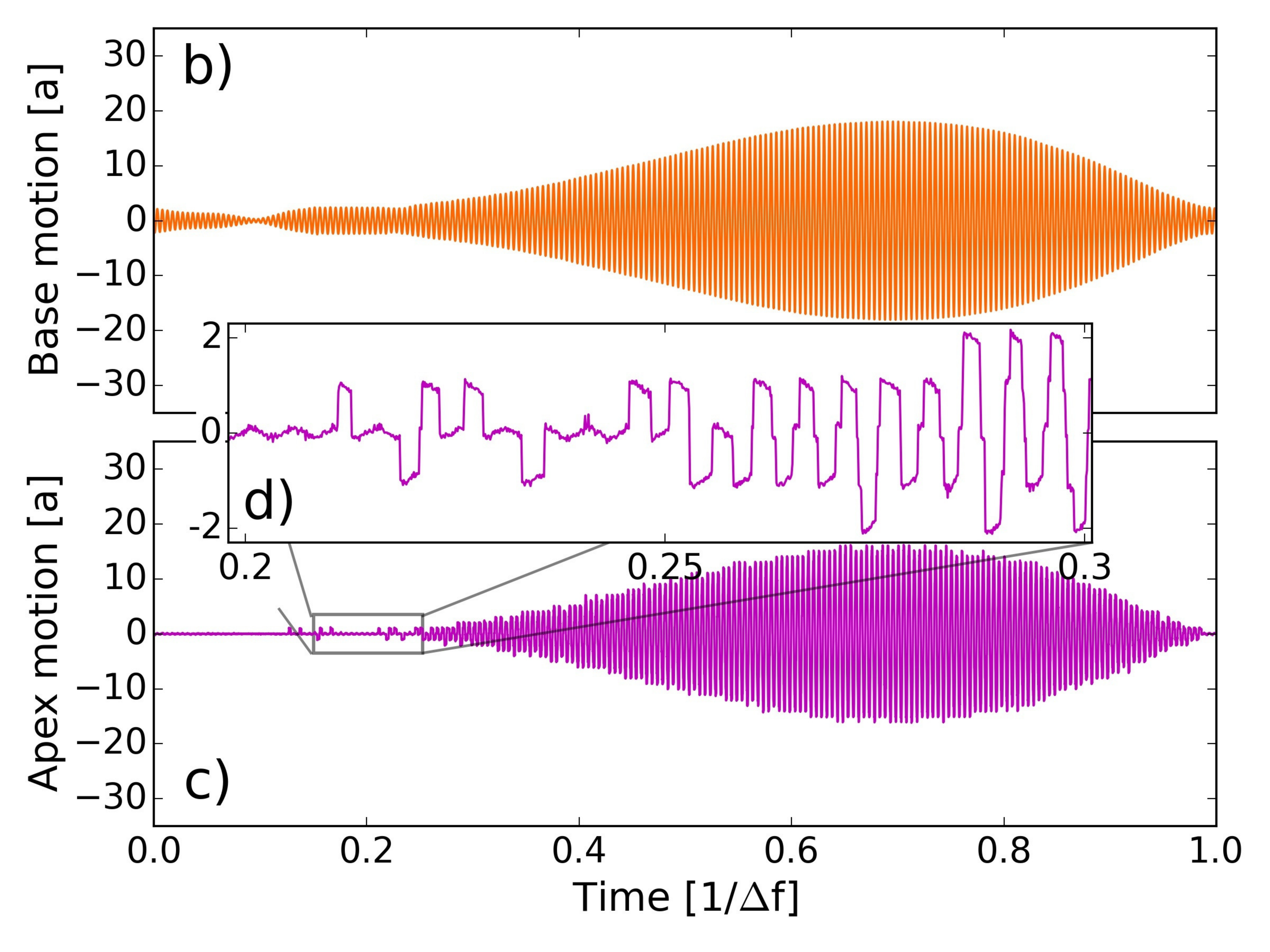}
\caption{\textbf{Simulated response.} a) Simulated frequency domain response of the tip base $x_\mathrm{b}$. The discrete comb of response at intermodulation frequencies $f_\text{IMP} = n_1f_1 + n_2f_2$ results from the periodic drive and the nonlinearity. b) and c) One period of steady state motion in the time domain, for both tip base $x_\mathrm{b}$ and the tip apex $x_\mathrm{t}$ when $U_0=2.2$, $\eta=7.6$ (orange curve in fig. \ref{fig:tap}). The elastic tip allows for motion of the base even when the apex is stuck to the surface.  The zoom inset d) shows the stick-slip region. 
}
\label{fig:ptsims}
\end{figure}

Figure~\ref{fig:tap} c) and d) show the simulated force quadratures (see Methods).
Adjusting the parameters of the asperity, we can achieve good qualitative agreement between the experimental and simulated curves. 
Simulation allows for detailed examination of the system dynamics during the transition from stick-slip to sliding friction.  In the frequency domain (figure \ref{fig:ptsims}a), the periodic motion of the tip base is represented by a frequency comb.  In the time domain (figures \ref{fig:ptsims}b and \ref{fig:ptsims}c) the motion of both the tip base and tip apex are plotted over exactly one period $T=1/\Delta f$, where $\Delta f=f_2-f_1$ is the frequency difference of the two drive tones.  

At low drive amplitude the tip apex becomes stuck in a local minimum of the potential.  The tip base continues to oscillate because the elastic tip can deform.  With increasing drive amplitude the tip apex begins to jump between local minimum of the potential as shown in fig.~\ref{fig:ptsims}~d).  When the drive amplitude is large enough stick-slip behavior gives way to smooth-sliding over many minima in the surface potential.  

\begin{figure}
\hspace*{-1.1cm}  
\includegraphics[width=\columnwidth]{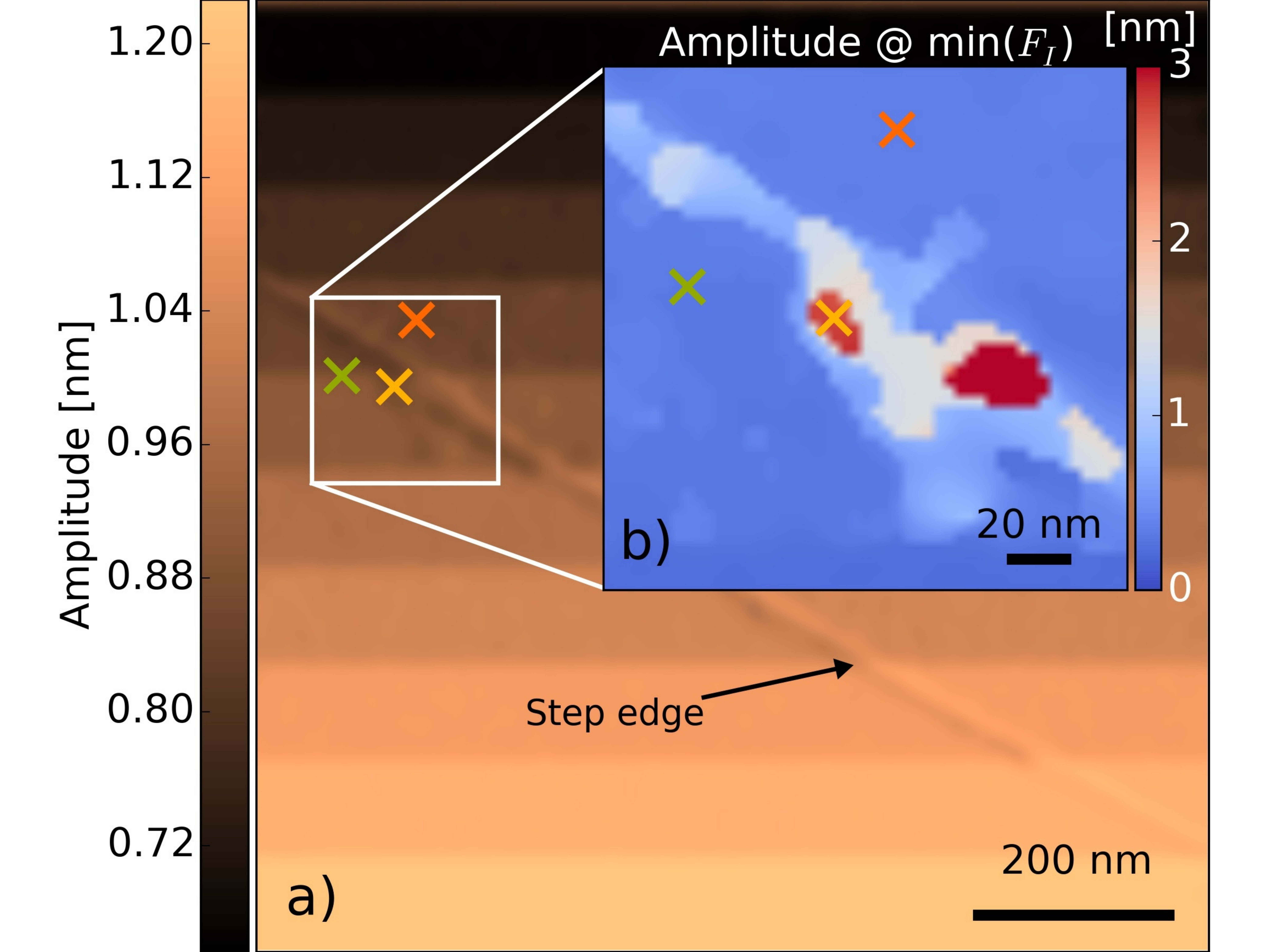}
\hspace*{0.1cm}  
\includegraphics[width=0.97\columnwidth]{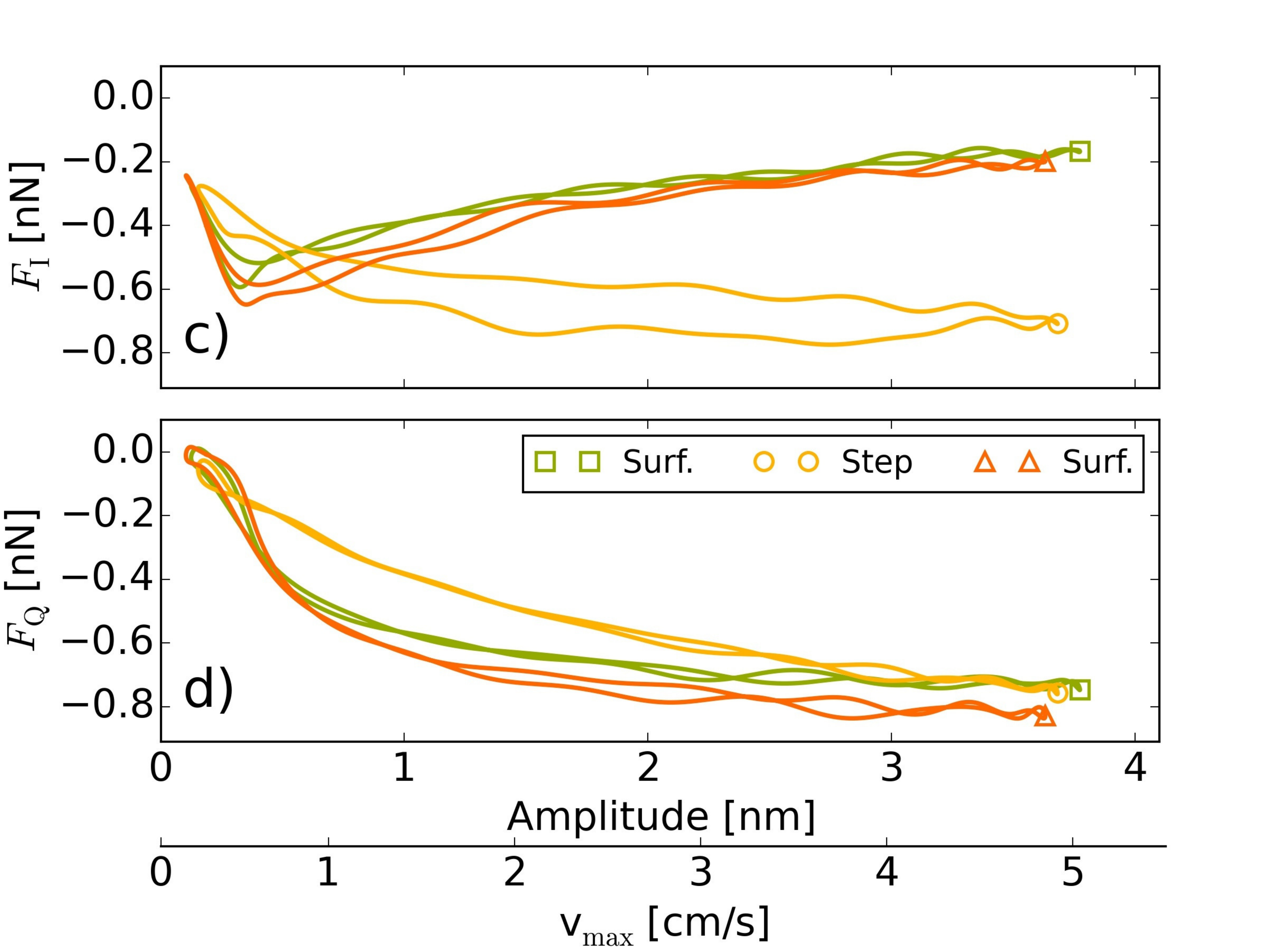}
\caption{\textbf{Friction images of a HOPG surface.}  a) The response amplitude at drive frequency $f_1$, used for scanning feedback.  The horizontal bands are due to changes in the feedback set-point during the scan, where lower amplitude (darker) corresponds to the cantilever working closer to the surface. The diagonal feature is an atomic step.  b) A zoom of the step region, where the image color codes for the amplitude at which $F_I$ is minimum.  c) and d) The force quadrature curves  at the three pixels marked with an $\times$ of corresponding color in the images.
}
\label{fig:image}
\end{figure}

\section*{Discussion}

The experimental curves figs. \ref{fig:tap} a) and b) show how the transition from stick-slip to smooth-sliding changes with applied load.  The simulations in figure \ref{fig:tap}~c) and d) capture the qualitative shape of the force quadrature curves at higher load force.  We find that the simulations become unstable at lower load, when the tip is just grazing the nonlinear surface interaction.  With sufficient interaction strength, the tip becomes stuck to the surface and the low amplitude slope of $F_I(A)$ gives the elastic stiffness of the tip $k$ (see Methods).  For this probe we measure $k=4$ N/m, consistent with estimates made by other groups on similar probes \cite{Krylov2006, Reimann2004}.  At lower load force a detailed examination of the experimental curves shows hysteresis in the force quadratures, as the low amplitude sticking regime ($F_Q=0$) gradually disappears with reducing load force. 

Intermodulation frictional force microscopy (ImFFM) breaks new ground with its unique ability to probe friction at high velocity with high spatial resolution.  Only 2~ms is needed to measure the force quadrature curves at the nN force scale and cm/s velocity scale.  This time is short enough to scan at a typical rate for dynamic AFM (1~line/s 256~pixels/line) and create an image of the transition from stick-slip to smooth sliding.  Figure \ref{fig:image} a) shows such a scan over a graphite surface, where the response amplitude at drive frequency $f_1$ is shown by color.  The feedback, which adjusts the probe height so as to keep this amplitude constant, was changed at regular intervals during the scan, resulting in the horizontal bands seen in the image.  Stable imaging was observed and there was no discernible evidence that the tip was damaging the surface, even at the highest load force.  

Graphite serves as a well-studied test sample for demonstration of ImFFM but the image is basically featureless because the friction is so homogeneous.  However, a change in the response is observed when scanning across an atomic step, seen as a diagonal feature in fig.~\ref{fig:image}~a).  The inset figure~\ref{fig:image}~b) shows a zoom of the step region where the color map codes for the critical oscillation amplitude at which $F_I(A)$ is minimum.  In this region three pixels are marked with a $\times$, and the $F_I$ and $F_Q$ curves are shown in figure \ref{fig:image} c) and d) with corresponding color.  Taking the minimum of $F_I(A)$ as the onset of sliding friction, one can see how the presence of the atomic step pushes the onset to larger amplitude.  The zoom, which is derived from all the amplitude and phase images measured at each of the intermodulation frequencies, shows features that are not present in any single image, and it demonstrates the remarkable detail with which high velocity friction can be studied using ImFFM.  The spatial resolution is limited only by the extent of the lateral tip oscillation, $2A \simeq 7.2$~nm for this scan.  With its high spatial resolution, and its ability to capture the full amplitude dependence of friction at each image point, we anticipate that ImFFM will have large impact on our understanding of the origins of friction on heterogeneous nanostructured surfaces.

\section{Methods}
\subsection*{Sample, cantilever and calibration}

We scanned a freshly cleaved highly oriented polylithic graphite (HOPG) sample under ambient conditions. The cantilever (MPP-13120 also known as Tap525, Bruker) was calibrated using the noninvasive thermal noise method developed for flexural eigenmodes \cite{Higgins2006}. From the thermal noise spectrum of the first flexural eigenmode \cite{Hutter1993} we determine the resonance frequency  $f_{0, \mathrm{f}} = 470$~kHz and quality factor $Q_\mathrm{f} = 384$.  The normal Sader method \cite{Green2004} is used to get the flexural stiffness $k_{\mathrm{f}}=53$~N/m.   Similarly, for the first torsional resonance $f_{0, \mathrm{t}} = 2400$~kHz , $Q_\mathrm{t} = 704$ and the torsional Sader method \cite{Green2004} gives a torsional stiffness $k_\phi = 239\cdot 10^{-9}$~Nm/rad. Together with the fluctuation-dissipation theorem we can get the detectors inverse responsivity $\alpha_t^{-1} = 1.2 \times 10^{3}$~rad/V.  This torsional stiffness corresponds to a stiffness for in-plane forces acting on the tip, $K = k_\phi/h_\mathrm{tip}^2 = 827$~N/m (manufacturer specified tip height  $h_\text{tip} = 17$~$\mu$m). We formulate the equations of motion below in terms of this equivalent lateral stiffness of the torsional eigenmode, with its associated mass $M= K/ (2 \pi f_{0, \mathrm{t}})^2$ and damping coefficient $M \Gamma =  K / 2 \pi f_{0, \mathrm{t}} Q_\mathrm{t}$, where $\Gamma$ is the width of the resonance.

\subsection*{Force sensitivity and image resolution}

The sensitivity of a cantilever as transducer of force is enhanced by a factor $Q$ on resonance, in comparison to the quasi-static (zero frequency) limit. Due to this enhancement the thermal Brownian motion of the cantilever can often be observed as a noise peak at resonance, where the Brownian motion noise exceeds the detector noise.  In this case the minimum detectable lateral force acting on the tip is given by the thermal noise force, with power spectral density,
\begin{equation}
S_{FF} = 2 k_\text{B} T M \Gamma = 2 k_\text{B} T \frac{k_\phi^2}{h_\text{tip}^2 2 \pi f_0 Q} \qquad \text{N$^2$/Hz}.
\end{equation}
Note that this noise force depends on the damping coefficient, not the stiffness, but it is convenient to express it in terms of stiffness, quality factor and resonant frequency, as the later two quantities are  easily accessible in the experiment.  For a specified measurement bandwidth $B$ (inverse of the measurement time), the minimum detectable force is the force signal which just equals this noise $F_\text{min} = \sqrt{S_{FF} B}$.  At the first torsional eigenmode of our cantilever with $B=1$~kHz, we find $F_\text{min} = 0.88$~pN.  

We compare with the quasi-static sensitivity where the measurement bandwidth is centered at zero frequency. Detector noise is typically limiting sensitivity with a noise equivalent force given by,
\begin{equation}
S_{FF}^\text{equiv} =    \frac{ S_{VV} k_\phi^2}{  \alpha_t^2  h_\text{tip}^2} \qquad \text{N$^2$/Hz}
\end{equation}
We take voltage noise $S_{VV} = 8.0 \times 10^{-12}$~V$^2$/Hz and inverse responsivity  $\alpha_t^{-1} = 1.2 \times 10^{-3}$~rad/V typical of our detector.  Quasi-static measurement typically uses a softer cantilever \cite{Alvarez-Asencio2013} $k_\phi \sim 3 \times 10^{-9} $~Nm/rad  which, for the same $h_\text{tip}=17$~$\mu$m and $B=$~1 kHz,  gives $F_\text{min}^\text{equiv} = 13$~pN, a factor of 15 less sensitive than our experiment. 

For quasi-static force measurement the time $1/B$ and constant sliding velocity $v$ determine the distance over which the force is measured, which defines a minimum feature size $\delta = v/B$.  Increasing the measurement bandwidth (decreasing the measurement time) improves resolution, but at the expense of force sensitivity.  With dynamic force measurement the minimum feature size is independent of the measurement bandwidth, given only by the amplitude of sliding oscillation $\delta=2A$, or in terms of the maximum velocity achieved in the oscillation $\delta = v_\text{max} /\pi f_0$.  High resolution (small $\delta$), high force sensitivity (small $F_\text{min}$) and high velocity (large $v_\text{max}$) are all achieved with a small bandwidth measurement on resonance using a cantilever with large $f_0$ and large $Q$. 

\subsection*{Intermodulation measurement and scanning feedback}

The cantilever is excited with a split-piezo actuator at two frequencies $f_1, f_2$ centered on torsional resonance $f_{0,\text{t}}$  and separated by $\Delta f = f_2 - f_1 \ll f_{0,\text{t}}$.  The drive frequencies $f_1$ and $f_2$ are chosen such that they are both integer multiples of $\Delta f$. The drive is synthesized, and the response is measured with a synchronous multifrequency lockin amplifier (Intermodulation Products AB) \cite{imp, Tholen2011}, which also calculates the feedback error signal used by the host AFM.  A proportional-integral feedback loop adjusts the probe height so as to keep the $f_1$ response amplitude at the set-point value.  The exact type of feedback used is not critical to the method, only that it is responsive enough to track the surface topography at the desired scan speed.  We also desire that the feedback error is small enough, such that we can approximate the probe height as being constant during the time $T=1/\Delta f$ needed to measure the response.  This time defines one pixel of the 42 amplitude and phase image-pairs acquired at each frequency, during a single scan. 

\subsection*{Model and equations of motion~\label{sec:pt}}

A schematic representation of the model can be seen in fig. \ref{fig:twist}. Performing force balance on both masses results in two coupled one-dimensional equations of motion in the lateral position of the tip apex $x_\mathrm{t}$, and tip base $x_\mathrm{b}$. 
\begin{eqnarray}
M\ddot{x}_\mathrm{b} &=& -Kx_\mathrm{b} - \Gamma M \dot{x}_\mathrm{b} + F_\text{c}(d, \dot{d}) + F_\text{d}(t),
\label{eq:model1}
\\
m\ddot{x}_\mathrm{t} &=& - F_\text{c}(d, \dot{d}) - F_\text{surf}(x_\mathrm{t}, \dot{x}_\mathrm{t}) 
\label{eq:model2}
\end{eqnarray}
The coupling force $F_\text{c}= k d+ m \gamma \dot{d} + F_\mathrm{noise}$ is linear in the deformation of the tip, $d = x_\mathrm{t} - x_\mathrm{b}$, and damping linear in $\dot{d}$.  $F_\mathrm{noise}(t)$ is a random noise force with a Gaussian distribution \cite{Vanossi2013}.  The strength of the noise is characterized by the standard deviation $\sigma_\mathrm{noise}$, given in table I. The nonlinear frictional force $F_\text{surf} = - \eta \dot{x}_\mathrm{t} - \frac{\partial}{\partial x_\mathrm{t}}U(x_\mathrm{t})$ is derived from damped motion in a periodic potential $U(x_\mathrm{t})=U_0 \cos(2 \pi x_\mathrm{t} /a_0)$.  The drive force $F_\text{d} = K\left[A_1\cos(2\pi f_1 t) + A_2\cos(2\pi f_2 t)\right]$ is applied at two frequencies as described above.

\subsection*{Dynamic force quadratures}

We probe friction by measuring two dynamic quadratures of the nonlinear force which is perturbing the harmonic motion of the torsional resonance.  The method was originally developed for normal forces and flexural resonance by Platz {\em et al.}\citep{Platz2012, Platz2013}.  From the measured intermodulation spectrum and the calibrated transfer function of the torsional eigenmode, we determine the oscillation amplitude-dependence of the force quadratures, without any assumptions as to the nature of the perturbing force.  For the model described above, $F_I$ gives the integrated coupling force $F_\text{c}$ that is in phase with the motion of the tip base, and $F_Q$ that which is quadrature to the motion, or in phase with the velocity.
\begin{equation}
F_I(A) = \frac{1}{T}\int_0^T F_\text{c}(x_\mathrm{b}, \dot{x}_\mathrm{b})\cos(\omega_0 t)dt,
\label{eq:fi}
\end{equation}
\begin{equation}
F_Q(A) = \frac{1}{T}\int_0^T F_\text{c}(x_\mathrm{b}, \dot{x}_\mathrm{b})\sin(\omega_0 t)dt,
\label{eq:fq}
\end{equation}
where  
\begin{equation}
x_\mathrm{b}(t)=A\cos(\omega_0 t) 
\end{equation}

When  $F_\text{fric} \gg F_{c}$, the tip apex is stuck in a minimum of the surface potential, $x_\mathrm{t} \approx \text{const}$, and motion of the tip base is due to tip deformation alone.  In this case we can solve the integrals in Eqs.~\eqref{eq:fi} and \eqref{eq:fq},
\begin{equation}
F_\text{I}(A) = -\frac{kA}{2}\text{ and } F_\text{Q} = -\frac{m\gamma v_\text{max}}{2}.
\label{eq:tip_stiffness}
\end{equation}
Thus, the slope of $F_I(A)$  at low amplitude and high load gives the stiffness of the asperity.  Similarly, the slope of $F_Q(A)$ gives the damping of the asperity, which is not resolvable in our experiment.

\subsection*{Simulation}

We simulate the experiment by numerical integration of the model Eqs.~(\ref{eq:model1}) and (\ref{eq:model2}) using CVODE\cite{SUNDIALS}.  The dynamical system is converted to 4 first-order differential equations, characterized by two resonant frequencies: $\omega_{0,b} = \sqrt{K/M}$ and $\omega_{0,t} = \sqrt{k/m}$.  When $\omega_{0,t} \gg \omega_{0,b}$ the adaptive time-step integrator becomes rather slow.  We chose  $\omega_0^t / \omega_0^b \sim 300$, which is at least one order of magnitude smaller than experiments, 
but still large enough to simulate the dynamics qualitatively so that we can explore the parameter space of the model in a reasonable time (each simulation takes 200 sec. on an Intel Core i7, 3.50GHz PC).
We simulated with normalized values where $\mathrm{[length]}=1.42$\AA, $\mathrm{[mass]}=4.78\cdot 10^{-25}$kg and $\mathrm{[time]} = 4.46 \cdot 10^{-13}$s.  The parameters are given in Table I and II.
\begin{table}
    \begin{tabular}{| c | c | c | c |}
    \hline
    Symbol & Expression & Value & Description \\ \hline
    m & - & 2.5 & Tip mass \\ \hline
    k & - & 3.6 & Tip spring \\ \hline
    $\gamma$ & - & 12 & Tip damping \\ \hline 
    \hline
    K & - & 40 &  Cantilever spring \\ \hline
    f$_0$ & $\sqrt{K/M}/2\pi$ & 0.001 &  Cantilever resonance \\ \hline
    Q & $2\pi f_0/\Gamma$ & 500 &  Cantilever quality factor \\ \hline
    \hline
    $\Delta\mathrm{f}$ & $f_0/200.5$ & - &  Frequency spacing \\ \hline
    $f_1$ & $200\Delta \mathrm{f}$ & - &  First drive frequency \\ \hline
    $f_2$ & $201\Delta \mathrm{f}$ & - &  Second drive frequency \\ \hline
    $A_1$ & - & 0.21 &  First drive amplitude \\ \hline
    $A_2$ & - & 0.21 &  Second drive amplitude \\ \hline
    \hline
    $a_0$ & - & 1 &  Surface periodicity \\ \hline
    \hline
    $\sigma_\mathrm{noise}$ & - & 6.1 &  Strength of noise \\ \hline
    \end{tabular}
    \caption{Parameters used for simulation of the PT model.  These parameters are kept constant.}
\end{table}
To simulate different interaction strengths, we vary the surface potential $U_0$ and dissipation of the tip $\gamma$ as in table 2. 
\begin{table}
    \begin{tabular}{| c | c | c | c |}
    \hline
    $U_0$ & $\eta$ & Symbol in fig 2b) & Description \\ \hline
    2.40 & 7.81 & Circle & Strongest interaction \\ \hline
    2.16 & 7.56 & Square & $\downarrow$ \\ \hline
    1.92 & 6.55 & Triangle &  Weakest interaction \\ \hline
    \end{tabular}
    \caption{Parameters to simulate different interaction force.}
\end{table}
Our choice of simulation parameters means that the simulated frequency of surface-induced force pulses on the tip $f_\text{surf} \sim (A/a_0) f_{0,t}$ is about an order of magnitude smaller than in the experiment.  
Nevertheless, our simulation is able to capture the qualitative shape of the force quadrature curves at high velocity and high interaction strength.  However, with these simulation parameters we are not able to reproduce the experiment at low velocity and low interaction.

\section*{Acknowledgements}

We gratefully acknowledge financial support from the Swedish Research Council (VR), the Knut and Alice Wallenberg Foundation, and the Olle Engkvist Foundation.  We also acknowledgement the use of methods and analysis code originally developed by Daniel Platz, as well as fruitful discussions with Mark Rutland and Roland Bennewitz.

\section*{Author contribution}

All authors contributed to discussion and interpretation of the experimental data, model and simulations.  PAT did the measurements and data analysis, performed the simulations and  generated all figures.  ASdW contributed with model development and simulation code.  RB and DF contributed to the experiments and simulation code.  DBH, ASdW and PAT contributed to the writing of the manuscript.

\section*{Competing financial interests}

DBH and DF are part owners of the company Intermodulation Products AB, which manufactures and sells the multifrequency AFM system used in this work.  

\end{document}